# BENDING LOSS REGULARIZED NETWORK FOR NUCLEI SEGMENTATION IN HISTOPATHOLOGY IMAGES

*Haotian Wang, Min Xian\*, Aleksandar Vakanski*

Department of Computer Science, University of Idaho, Idaho, USA

## ABSTRACT

Separating overlapped nuclei is a major challenge in histopathology image analysis. Recently published approaches have achieved promising overall performance on public datasets; however, their performance in segmenting overlapped nuclei are limited. To address the issue, we propose the bending loss regularized network for nuclei segmentation. The proposed bending loss defines high penalties to contour points with large curvatures, and applies small penalties to contour points with small curvature. Minimizing the bending loss can avoid generating contours that encompass multiple nuclei. The proposed approach is validated on the MoNuSeg dataset using five quantitative metrics. It outperforms six state-of-the-art approaches on the following metrics: Aggregate Jaccard Index, Dice, Recognition Quality, and Panoptic Quality.

*Index Terms—* Nuclei segmentation, histopathology images, bending loss, multitask deep learning

## 1. INTRODUCTION

Histopathology image analysis provides direct and reliable evidence for cancer detection. Conventionally, pathologists examine the shapes and distributions of the nuclei under microscope to determine the carcinoma and the malignancy level [1]. The large amount of nuclei makes the whole process time-consuming, low-throughput, and prone to human error. Recently, many computational approaches have been proposed for automatic nuclei segmentation. However, it is still challenging to segment nuclei in histopathology images accurately due to color and contrast variations, background clutter, image artifacts, and large morphological variances.

To overcome the challenges, some approaches [2-4] utilized thresholding and watershed algorithms to segment nuclei, but these approaches are not robust in handling images with various nuclei types, fat tissue, and staining procedure. In recent time, deep learning-based approaches have been thriving in numerous biomedical image processing tasks [5-7], and have achieved promising results in nuclei segmentation [8-13]. Kumar et al. [8] demonstrated a 3-class (instance, boundary, and background) convolutional neural network (CNN) to segment overlapped nuclei. The approach computes the label for each pixel, which is slow and sensitive to noise. Chen et al. [9] proposed a multitask learning framework that outputs both an instance map and a boundary map in separate branches. Naylor et al. [10] constructed a regression network that generated markers for watershed algorithm to segment overlapped nuclei. Graham et al. [11] proposed a new weighted cross-entropy loss that is sensitive to the Hematoxylin stain for nuclei segmentation. Oda et al. [12] proposed a new CNN architecture with two decoding paths; one path was designed to enhance the boundaries of cells; the other path was to enhance the entire instance segmentation. Zhou et al. [13] proposed the CIA-Net that utilizes spatial and texture dependencies between nuclei and contours to improve the robustness of nuclei segmentation. Most approaches focused on developing more complex and deeper neural networks. Based on the reported results, they achieved better overall performance than traditional methods, but their ability to separate overlapped nuclei still is limited (Fig. 1).

\* Correspondence to Min Xian (mxian@uidaho.edu). This work was supported, in part, by the Center for Modeling Complex Interactions (CMCI) at the University of Idaho through NIH Award #P20GM104420.

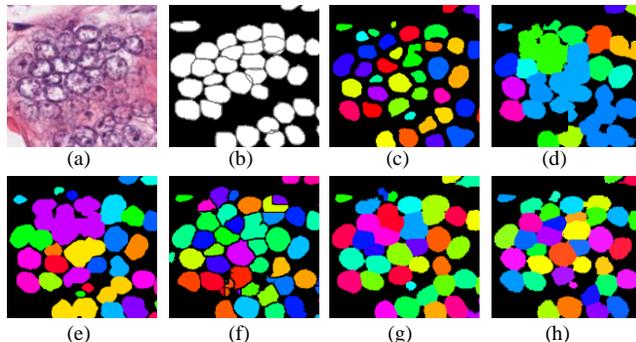

**Fig. 1.** Examples of state-of-the-art approaches in segmenting touching nuclei. (a) image patch; (b) ground truth; (c) U-Net [6]; (d) SegNet[7]; (e) DCAN [9]; (f) DIST [10]; (g) HoVer-Net[22]; and (h) ours.

To solve the challenge, we proposed a new bending loss and integrated it as a regularizer into a deep multitask learning framework. Naturally, in histopathology images, the curvatures of a nucleus contour points change smoothly; but, if one contour contains two or multiple overlapped or touching nuclei, their touching points on the contour will have large curvature changes (Fig. 2). Inspired by this observation, we proposed the bending loss to penalize the curvature of contour points. Specifically, the bending loss results in a small value for a contour point with small curvature; and generates a large penalty for a point with large curvature. Minimizing the bending loss will produce nuclei boundary points with smooth curves; consequently, it can avoid generating boundaries for two or multiple touching nuclei.

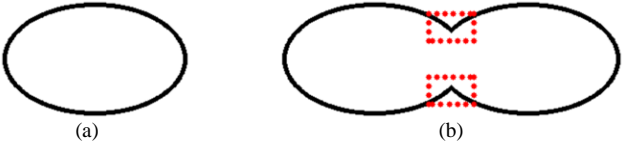

**Fig. 2.** An example of nuclei contours. (a) A nucleus contour; and (b) a contour contains two nuclei. Red rectangles highlight the touching points on the contour.

## 2. THE PROPOSED METHOD

### 2.1. Bending Loss Regularized Network

Bending energy has been widely applied in measuring the shapes of biological structures, e.g., blood cells [14], cardiac [15], vesicle membranes [16], and blood vessels [17]. Young et al. [18] used the chain-code representations to model bending energy. Vliet et al. [19] used the derivative-of-Gaussian filter to model bending energy in gray-scale image for motion tracking. Wardetzky et al. [20] modeled the discrete curvature and bending loss both in kinematic and dynamical treatment to solve the smoothness problem. Inspired by [20], we proposed a rotation-invariant bending loss for penalizing the curvature changes of nuclei contours. The proposed method calculates bending loss for every point on a contour, and defines the bending loss of an image patch as the mean discrete bending loss of all contours points.

For 2D digital images, a contour is composed of discrete pixels, and the curvature of a specific contour point is computed by using the vectors created by neighboring points on the contour. For histopathology images, a nucleus usually has a smooth contour, and the points on the contour have small curvature changes; the points on the contour with large curvature changes have high probability to be the touching points of two/multiple nuclei. To split the touching nuclei, we can define the bending loss that gives high penalties to the contour points with large curvature, and small penalties to points with small curvature. The proposed loss function is given by

$$L = L_1 + \alpha \cdot L_{Bend} \quad (1)$$

where $L_1$ refers to the conventional loss for image segmentation, e.g., cross-entropy, dice loss, or mean squared error; $L_{Bend}$ denotes the proposed bending loss; the parameter $\alpha$ controls the contribution of the bending loss. Let $B$ be all contour points of nuclei in an image, and the bending loss of the image is defined by

$$L_{Bend}(B) = \frac{1}{m}\sum_{i=1}^{m} BE(i) \quad (2)$$

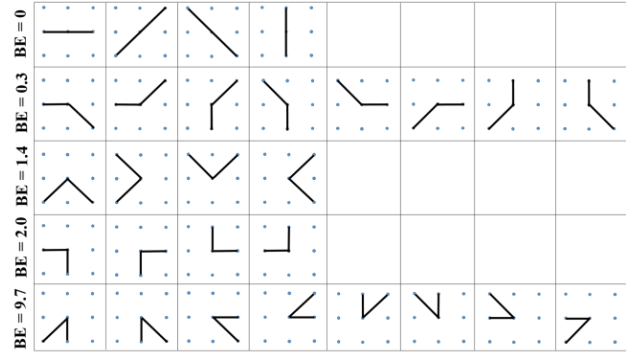

**Fig. 3.** Discrete bending losses for different curve patterns.

where the contour thickness is one pixel, $m$ is the number of contour points, and $BE(i)$ is the discrete bending loss for the $i$th point given by

$$BE(i) = \frac{\kappa(i)^2}{|v(i,i+1)| + |v(i-1,i)|} \quad (3)$$

$$\kappa(i) = \frac{2|v(i-1,i) \times v(i,i+1)|}{|v(i-1,i)||v(i-1,i)| + v(i-1,i) \cdot v(i,i+1)} \quad (4)$$

where $\kappa(i)$ is the curvature at the $i$th point; for three consecutive pixels on a nucleus boundary with coordinates $x_{i-1}$, $x_i$ and $x_{i+1}$, $v(i-1, i)$ is the edge vector from point $i-1$ to $i$, such that $v(i-1, i) = x_i - x_{i-1}$; and $v(i, i+1)$ is the edge vector from $i$ to $i+1$, such that $v(i, i+1) = x_{i+1} - x_i$; and operator $|\cdot|$ denotes the length of a vector.

The 8-neighborhood system is applied to search neighbors for contour points. Ideally, a contour point only has two neighboring points, and their coordinates are used to calculate the edge vectors in Eqs. (2) and (3). As shown in Fig. 3, a point with eight neighbors has 28 combinations of possible contour shapes. All shape patterns are divided into five groups, and in each group all shapes have the same bending loss. In the first row, the four patterns construct straight lines, and their bending loss are all 0s. The second row shows patterns with $3\pi/4$ angle between edge vectors, and their bending loss are small (0.3). In the last row, the eight patterns have large curvatures, and their bending loss are the largest in all patterns. The third and fourth rows illustrate patterns with same angles between edge vectors; they have different bending loss due to the difference of the vector length.

The proposed bending loss is rotation invariant since all patterns with the same angle between edge vectors have the same bending loss. In practice, if two nuclei contours share some contour segments, one contour point may have more than two neighbors. In this scenario, we calculate the bending loss for all possible combinations, and choose the smallest loss as the discrete bending loss for the current contour point.

The 8-neighborhood system is chosen by experiment. We found that the nuclei touching points have remarkably large curvature change, which is significantly larger than the curvatures of the contour points on well-segmented nuclei.

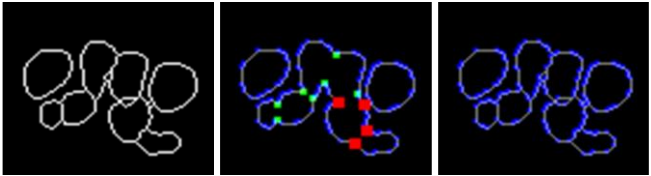

(a) Nuclei contours    (b) Poorly-segmented    (c) Well-segmented

**Fig. 4.** Bending loss examples. (a) Ground truth of eight nuclei contours; (b) BE values of contour points of poorly-segmented nuclei; and (c) BE values of well-segmented nuclei. Red: BE = 9.7, green: BE = 1.4 and BE = 2.0, blue: BE = 0.3, and grey: BE = 0.

Also, for other larger neighborhood systems, large bending loss could be generated from a smooth contour.

As shown in Fig. 4, for poorly-segmented nuclei contours, all these touching points have relatively high (red and green points) bending loss. In Fig. 4(c), the touching nuclei are well separated, and the bending loss of all contour points are less than 0.3.

### 2.2. Nuclei Segmentation Scheme

The deep learning scheme comprises three stages: 1) preprocessing; 2) bending loss regularized network; and 3) post processing. The preprocessing performs color normalization [21] to reduce the impact of variations in the H&E staining and scanning processes. The postprocessing described in the multitask learning [22] is employed in this study. It followed the encoder-decoder architecture, and used a pretrained 50-layer ResNet [23] as the encoder. The decoder contains two branches: nuclei instance branch and distance map branch. The nuclei instance branch predicts the inner regions of the nuclei, whereas the distance map branch outputs horizontal and vertical distances of nuclear pixels to their centers of mass. The loss function (Eq. 1) consists of $L_1$ term from [22] and the weighted bending loss term defined in Eq. (2). Minimizing the whole loss will enforce the network to output contours with smaller curvatures. The hyperparameter α is set by empirical experimentaion. The postprocessing first applies Sobel operators to the distance maps generating initial nuclei contours; then the initial contours are combined with the nuclei instance map to produce watershed markers, and finally the watershed algorithm is applied to generate nuclei regions.

## 3. EXPERIMENTS AND RESULTS

### 3.1. Dataset and Evaluation Metrics

The proposed method is evaluated using the MoNuSeg image set [8] which contains 30 images from TCGA (The Cancer Genomic Atlas) dataset. The original size of the images is $1000 \times 1000$ pixels, and the nuclei instances (21,000) were manually annotated. These images covered seven different organs (breast, liver, kidney, prostate, bladder, colon, and stomach). In the experiment, 16 images (4 breasts, 4 livers, 4 kidneys, 4 prostates) are used for training and validation, and 14 images for test. The training and validation sets contain over 13,000 annotated nuclei; the test set has 6,000 nuclei, and is split into the *same organ* set (2 breasts, 2 livers, 2 kidneys, and 2 prostates) and *different organ* set (2 bladders, 2 colons, and 2 stomachs). The input image size to the network is 270×270×3. We prepare our training, validation, and test sets by extracting image patches from images with 270×270 pixels size. During the training stage, data augmentation strategies, i.e., rotation, Gaussian blur, and median blur are utilized for generating more images; hence, the final training and validation sets contain 1,936 images.

*Evaluation metrics.* We employed five metrics to evaluate the performance of nuclei segmentation approaches: Aggregate Jaccard Index (AJI) [8], Dice coefficient [24], Recognition Quality (RQ) [25], Segmentation Quality (SQ) [25], and Panoptic Quality (PQ) [25]. The performance of six recently published approaches [5-7, 9-10, 22] is evaluated and compared with the proposed approach. Let $G = \{G_i\}_{i=1}^{N}$ be the ground truths segments in an image, $N$ denotes the total amount of segments in $G$; and let $S = \{S_k\}_{k=1}^{M}$ be the predicted segments of the corresponding image, $M$ denotes the total amount of segments in $S$. AJI is an aggregate version of Jaccard Index and is defined by

$$\text{AJI} = \frac{\sum_{i=1}^{N} G_i \cap S_j}{\sum_{i=1}^{N} G_i \cup S_j + \sum_{S_k \in U} S_k} \quad (5)$$

where $S_j$ is the matched predicted segments that produce the largest Jaccard Index value with $G_i$; and $U$ denotes the set of unmatched predicted segments, where the total amount of $U$ is $(M - N)$. Dice coefficient (DICE) is utilized to evaluate segmentation results, DICE is given by

$$DICE = \frac{2|G \cap S|}{(|G| + |S|)} \quad (6)$$

where operator $|\cdot|$ denotes the cardinalities of the segments.

Panoptic Quality (PQ) is used for quality estimation for both detection and segmentation problems. Recognition Quality (RQ) is a version of the familiar F1-score, and Segmentation Quality (SQ) is well-known as average Jaccard Index. *RQ, SQ, PQ* are defined as

$$RQ = \frac{TP}{TP + \frac{1}{2}FP + \frac{1}{2}FN} \quad (7)$$

$$SQ = \frac{\sum_{(p,g) \in TP} IoU(p,g)}{TP} \quad (8)$$

$$PQ = RQ \times SQ \quad (9)$$

where $p$ refers to predicted segments, $g$ refers to ground truth segments. The matched pairs $(p,g)$ are mathematically proven to be *unique matching* [25] if their IOU$(p,g) > 0.5$. The *unique matching* splits the predicted and ground truth segments into three sets: true positive (*TP*) is the number of matched pairs $(p,g)$, false positive (*FP*) is the number of unmatched predicted segments, and false negatives (*FN*) is the number of unmatched ground truth segments.

*Implementation and training*. The deep neural network is trained by using a NVIDIA GeForce GTX 1080 Ti GPU.

Table. 1. Overall Test Performance on the MoNuSeg Dataset.

|  | Same organ test | | | | | Different organ test | | | | |
|---|---|---|---|---|---|---|---|---|---|---|
|  | AJI | Dice | RQ | SQ | PQ | AJI | Dice | RQ | SQ | PQ |
| FCN8 [5] | 0.406 | 0.767 | 0.601 | 0.705 | 0.425 | 0.452 | 0.795 | 0.581 | 0.713 | 0.416 |
| U-Net [6] | 0.531 | 0.726 | 0.641 | 0.680 | 0.440 | 0.513 | 0.720 | 0.580 | 0.666 | 0.387 |
| SegNet [7] | 0.508 | 0.785 | 0.688 | 0.733 | 0.506 | 0.505 | 0.814 | 0.650 | 0.752 | 0.492 |
| DCAN [9] | 0.513 | 0.770 | 0.669 | 0.712 | 0.476 | 0.518 | 0.789 | 0.646 | 0.725 | 0.469 |
| DIST[10] | 0.563 | 0.786 | 0.747 | 0.757 | 0.566 | 0.593 | 0.825 | 0.708 | **0.780** | 0.554 |
| HoVer-Net [22] | 0.607 | 0.802 | 0.769 | 0.761 | 0.587 | 0.625 | 0.831 | 0.733 | 0.774 | 0.571 |
| Ours | **0.621** | **0.813** | **0.781** | **0.762** | **0.596** | **0.641** | **0.837** | **0.760** | 0.775 | **0.592** |

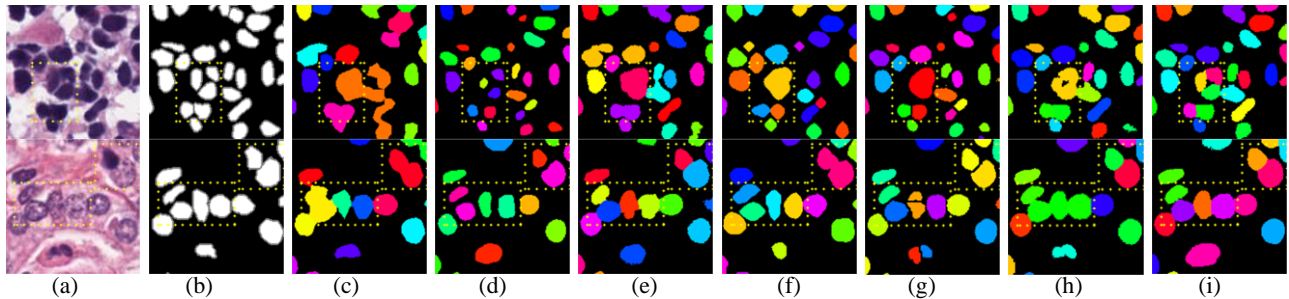

(a) (b) (c) (d) (e) (f) (g) (h) (i)

**Fig. 5.** Examples of segmentation results for touching nuclei. Image patch from left to right: (a) Original images, (b) ground truth, (c) FCN8, (d) U-Net, (e) SegNet, (f) DCAN, (g) DIST, (h) HoVer-Net and (i) ours.

The encoder was pretrained on ImageNet, then we trained the decoder using the same dataset for 100 epochs to obtain the initial parameters for the two decoder branches. The whole network was further fine-tuned for 100 epochs on our nuclei training set. The size of the final output images is 270×270 pixels, and these output images are merged to form images with the same size (1000×1000) as the original images. The initial learning rate is $10^{-4}$ and is reduced to $10^{-5}$ after 50 epochs, whereas the weight parameter of the bending loss α in the total loss function is set to be 1. The batch size is 8 for training the decoder and 2 for fine-tuning the network. Moreover, processing an image of size 1000×1000 with our architecture on average takes about three seconds.

### 3.2. Experimental Result

In this section, we compared the proposed approach to six recently published approaches: FCN8 [5], U-Net [6], SegNet [7], DCAN [9], DIST [10], and HoVer-net [22], using AJI, Dice, RQ, SQ, and PQ scores. Table.1 shows the overall results. Note that all other approaches are tested using the described experiment settings, and therefore, the values in Table 1 may not be the same as those reported in the original publications. The watershed algorithm is applied to FCN8, U-Net, and SegNet for postprocessing, whereas the rest of the approaches are implemented by following the same strategy as in the original paper.

As shown in **Table 1**, the proposed approach outperforms the other six approaches in overall performance. It achieves the highest AJI, Dice, RQ, SQ, and PQ values on same organ test and reaches the highest AJI, Dice, RQ, and PQ values on different organ test. The SQ score (0.775) in different organ test is also close to the highest score (0.780).

The bending loss is the major difference between the HoVer-Net and the proposed approach. The proposed approach outperforms HoVer-Net in all metrics, which proves the efficacy of the bending loss.

**Fig. 5** shows the segmentation results of several image patches with many touching nuclei. Each nuclei instance is represented with the same color, and the yellow rectangles highlight the overlapped nuclei in the image patches. The proposed approach and U-Net accurately segment all the overlapped nuclei. However, the predicted segments by U-Net are small, which easily cause over-segmentation problem, and leads to lower overall performance. From the results, it can be concluded that the proposed method achieved a better performance to segment and locate overlapped or touching nuclei compared with HoVer-Net. The results validation the assumption that our method can enhance the performance of the model for segmenting overlapped and touching nuclei.

## 4. CONCLUSION

In this paper, we proposed a bending loss regularized network to tackle the challenge of segmenting overlapped nuclei in histopathology images. The proposed method uses the nuclei curvature to define high penalties for touching points of overlapped contour segments and assigns small penalties to well-defined nuclei contours. Our method achieves the highest overall performance when compared to six other deep learning approaches on a public dataset. The proposed bending losss can be applied to other deep learning-based segmentation tasks. In the future, we will study the performance of the bending loss when applied at different image scales.